\documentclass[showpacs,aps,superscriptaddress,prb, twocolumn ,amsmath]
{revtex4}
\usepackage{amssymb}
\usepackage{graphicx,subfigure}
\usepackage[percent]{overpic}
\usepackage{dcolumn}
\usepackage{bm}
\usepackage{color}
\usepackage{hyperref}

\begin{document}

\preprint{APS/123-QED}

\title{3\emph{d} Transition Metal Adsorption Induced Vally-polarized Anomalous Hall Effect in Germanene}
\author{P. Zhou}
 \affiliation{Hunan Provincial Key laboratory of Thin Film Materials
and Devices, School of Material Sciences and Engineering, Xiangtan
University, Xiangtan 411105, China}
\affiliation{Key Laboratory of Low-dimensional Materials and Application Technology,  School of Material Sciences and Engineering, Xiangtan University, Xiangtan 411105, China}
\author{L. Z. Sun}
 \email{lzsun@xtu.edu.cn}
 \affiliation{Hunan Provincial Key laboratory of Thin Film Materials
and Devices, School of Material Sciences and Engineering, Xiangtan
University, Xiangtan 411105, China}
\affiliation{Key Laboratory of Low-dimensional Materials and Application Technology,  School of Material Sciences and Engineering, Xiangtan University, Xiangtan 411105, China}
\date{\today}
\begin{abstract}
Based on DFT+U and Berry curvature calculations, we study the electronic structures and topological properties of 3\emph{d} transition metal (TM) atom (from Ti to Co) adsorbed germanene (TM-germanene). We find that valley-polarized anomalous hall effect (VAHE) can be realized in germanene by adsorbing Cr, Mn, or Co atom on its surface. A finite valley hall voltage can be easily detected in its nanoribbon, which is important for valleytronics devices. Moreover, different valley-polarized current and even reversible valley Hall voltage can be archived by shifting the Fermi energy of the systems. Such versatile features of the systems show potential in next generation electronics devices.\\
\end{abstract}
 \maketitle
\section{INTRODUCTION}
\indent Besides charge and spin degree of freedoms, valley degree of freedoms of electrons in two dimensional (2D) crystals with honeycomb lattice structure has attracted intense interest recently due to its potential in next generation electronics devices, now called as valleytronics\cite{valley1,valley2,valley3,valley4,valley5}. The valley index can be regarded as a discrete degree of freedom for low-energy carriers, which is robust against smooth deformation and low-energy phonon thanks to the large separation of the valleys in momentum space. To utilize the valley index as an information for carrier, generation, detection, and identification of pure valley current and its response to external stimuli are therefore crucial premise.\\
\indent Graphene provides an excellent platform to make use of this freedom\cite{graphene1}. Valley-polarized phases can be acquired by breaking its space inversion symmetry. Although massive theoretical works have proposed the valley-polarized phenomenon in graphene, it is hard to realize in experiments due to its flat structure and weak spin-orbit coupling of C atom. This predicament has motivated researchers to investigate two-dimensionally ordered and layered materials analogous to graphene, such as silicene or germanene. Silicene and germanene have recently attracted much attention both in theoretical and experimental fields\cite{siexp1,siexp2,geexp1,geexp2,sical1,sical2} due to their versatile properties. Different from flat graphene, both silicene and germanene share buckled hexagonal lattice. It is easier to destroy their inversion symmetry and produce valley polarized phenomenon in the materials.\\
\indent Using first-principles method, Cahangirov et al.\cite{cry1} show that germanium atoms can form two-dimensional low-buckled honeycomb structures, germanene. They also indicate that charge carriers of germanene behave as a massless Dirac fermion. Subsequent study\cite{cry2} indicated that quantum spin Hall effect can be realized in germanene. Latest theoretical research also reported that functionalized germanene behaves as large-gap two-dimensional topological insulators\cite{funcGe}. In experiments, Bianco et al.\cite{geexp3} indicated that millimeter-scale crystals of a hydrogen-terminated multilayered germanene can be obtained from chemical deintercalation of $CaGe_2$. However, its semiconductor properties with nearly 1.59eV bandgap at $\Gamma$ point make it unable to make full use of Dirac electron of single layer germanene. The synthesis of pristine germanene on the surface of Au and Pt\cite{geexp1,geexp2} was recently reported, which pave the way for further manifesting of the valley polarization in the 2D germanene.\\
\indent Lately, valley-polarized quantum anomalous Hall effect (QAHE) was reported in silicene within tight-binding frame\cite{VAHE}. Anomalous Hall effect (AHE) is represented by anomalous Hall conductance occurring in magnetic materials\cite{qhermp}. Latest calculation also predict quantum AHE can be realized in  Co-decorated silicene\cite{sical1}. The valley-polarized quantum anomalous Hall insulator possesses quantum valley Hall effect (QVHE) and QAHE at the same time. Namely, it possesses the non-dissipative anomalous hall transportation and an additional new valley freedom. This new phase broadens the application of valleytronics in next-generation electronics.\\
\indent Inspired by the advancement of both germanene and QVHE, in this work, we concentrate on how to adjust the valley-polarization of germanene by introducing 3\emph{d} transition metal (TM) atom on its surface. In comparison with previous reports, we find that appropriate Hubbard U is important to accurately describe the TM-germanene systems. With the help of first-principles Berry curvature calculations, we find that 3\emph{d} TM atom can break the inversion symmetry of germanene and induce valley-polarization when the spin-orbit coupling (SOC) is included. Valley-polarized AHE can be achieved when Cr, Mn, or Co is adsorbed on the surface of germanene. Moreover, quantum valley Hall effect can be realized in Mn-germanene system just through shifting the Fermi level to specific energy window.\\
\section{COMPUTATIONAL METHODOLOGY}
\indent The electronic structures of germanene adsorbed 3\emph{d} TM atom including Ti, V, Cr, Mn, Fe, and Co were studied with projector augmented wave\cite{PAW}(PAW) formalism implemented in the Vienna ab initio simulation package\cite{vasp1,vasp2}. General gradient approximation\cite{pbe} was used to describe the exchange and correlation energy in the Kohn-Sham equations. The plane-wave cutoff energy was set to be 500 eV and a vacuum space larger than 15 {\AA} was set to avoid the interaction between two adjacent images. The energy convergence criterion was set to $10^{-6}$ eV/unit cell. We performed the structure optimization using the conjugated gradient algorithm. All of the atoms were allowed to relax without symmetric restriction until atomic residual forces were smaller than $10^{-2}$ eV/{\AA}. To accurately describe the electronic structure around the Dirac point, we considered the semicore state, such as 3\emph{d} of Ge and 3\emph{p} of TM, as valence state in our calculations.\\
\begin{table*}
\caption{Results for single 3\emph{d} TM atom adsorbed germanene. The results contain the Hubbard U of 3\emph{d} orbits of TM (U), adsorption energy of 3\emph{d} TM ($E_b$), distance between the adatom to its nearest Ge atoms of sublattice A/B (dTM-$Ge_A$/$Ge_B$), magnetic moment per unit cell (M), and charge transfer from the TM to the germanene ($T_e$). }\label{tab1}
\begin{ruledtabular}
\begin{tabular}{ccccccc}
  TM     &  $U$(eV)   &$E_b$(eV)& d$TM-Ge_A$  &d$TM-Ge_B$ & M($\mu_B$) & $T_e$  \\
  Ti     &  4.666 &4.807& 2.323 & 2.571 & 2.67(1.13\cite{gecal1}) & -1.153 \\
  V      &  3.868 &4.274& 2.530 & 3.073 & 4.67(3.01\cite{gecal1}) & -0.790 \\
  Cr     &  6.360 &1.489& 2.508 & 3.036 & 2.84(4.00\cite{gecal1}) & -0.672 \\
  Mn     &  4.732 &3.821& 2.414 & 2.844 & 5.75(4.95\cite{gecal1}) & -0.626 \\
  Fe     &  6.319 &3.306& 2.408 & 2.831 & 3.04(3.24\cite{gecal1}) & -0.567 \\
  Co     &  5.924 &4.146& 2.321 & 3.530 & 2.03(1.10\cite{gecal1}) & -0.285 \\
\end{tabular}
\end{ruledtabular}
\end{table*}
\indent We adopted the GGA+U method in our calculations because the electronic correlation is critical to accurately describe the properties of 3\emph{d} transition metal. To determine the parameter U, we used the linear response approach introduced by Cococcioni\cite{addu} implemented in the PWSCF package\cite{qe}. In the present work the rotationally invariant DFT+U formalism proposed by Dudarev\cite{duda} was used, where only the value of $U_{eff}=U-J$ is meaningful instead of individual U and J. To evaluate charge transfer between TM adatom and germanene sheet, we adopted the Bader charge analysis method\cite{bader}. By comparing the valence electrons of TM adatom in TM-germanene with its free-standing state, the charge transfer between TM and germanene can be quantitatively determined.\\
\indent For the part of anomalous Hall conductivity (AHC) calculation, we use the berry curvature formula:
\begin{eqnarray}\label{equ3}
{\Omega _n}(\bm{k}) =  - \sum\limits_{n^{'} \ne n} \frac{{2{\mathop{\rm Im}\nolimits} \langle {\psi _{n\bm{k}}}|{v_x}|{\psi _{n^{'}\bm{k}}}\rangle \langle {\psi _{n^{'}\bm{k}}}|{v_y}|{\psi _{n\bm{k}}}\rangle }}{{{{({\varepsilon _{n^{'}}} - {\varepsilon _n})}^2}}}
\end{eqnarray}
\begin{eqnarray}\label{equ4}
\sigma _{xy}=- \frac{e^2}{\hbar}\sum\limits_n {\int_{BZ}}{\frac{d\bm{k}}{(2\pi)^3}}{f_n}(\bm{k}){\Omega _{n,z}(\bm{k})}
\end{eqnarray}
$\Omega _n$ is berry curvature, $v_{x(y)}$ represents velocity operator, $\varepsilon _{n(n^{'})}$ donates the energy of bands calculated in normal first-principles method. As we known there are three main mechanisms for AHE: Intrinsic, Skew-scattering, and Side-jump contribution. We only considered the intrinsic mechanism of the AHE because the topological properties the main topic in present work is mainly determined by the intrinsic one\cite{AHE}. SOC was used when AHC calculations were performed. To ensure the accuracy of our calculations, we calculate the AHE of Fe and FePt, the results agree well with previous reports\cite{wannierahe,FePt}. Since a dense $k$ point mesh is needed for obtaining accurate AHC, we use the Wannier interpolation to obtain a high precision berry curvature in reciprocal lattice. In the method, we implemented an adaptive mesh refinement scheme\cite{adaptk} in k-point space when the computed berry curvature exceeds a threshold value ${\Omega}_{cut}$. In this paper, the k-point mesh of 5 $\times$ 5 $\times$ 1 is used in VASP calculations, then we interpolate the berry curvature in 60 $\times$ 60 $\times$ 1 in reciprocal space. We find that the submesh with 5 $\times$ 5 $\times$ 1 and ${\Omega}_{cut}$ = 90 a.u. are enough to achieve the precision of 0.1 $(\Omega m)^{-1}$.\\
\begin{figure}
{\includegraphics[width=3.5in]{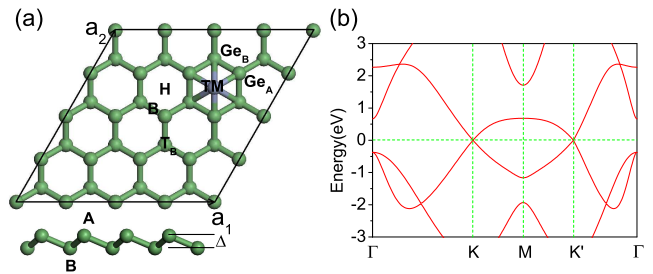}}\\
  \caption{(a) Top view of germanene monolayer where 3 adsorption sites (Hollow (H), top B sublattice ($T_B$) and Bridge (B)) are marked out with black letters. The lower panel of (a) is the side view of germanene, the two equivalent Ge sublattices are labeled as A and B, respectively, with a buckled distance $\Delta$. (b) The energy band of pristine germanene. Two equivalent valleys at $K$ and $K^{'}$ clearly locate around the Fermi level. }\label{FIG1}
\end{figure}
\indent To evaluate the relative stability of different adsorption sites, we calculate the adsorption energy of TM-germanene system defined as:
\begin{eqnarray}\label{equ3}
E_a = E_{TM+g} -E_{TM}-E_{g}
\end{eqnarray}
The terms $E_g$, $E_{TM}$, and $E_{TM+g}$ represent the total energies of the bare germanene, the free TM atom, and the TM-germanene system, respectively. The smaller $E_a$ means more stable structure. In present paper, 4$\times$4 supercells of germanene as shown in Fig.\ref{FIG1}(a) was employed to avoid the interaction of TM atoms.\\
\section{RESULATS AND DISCUSSIONS}
\indent Germanene consists of a honeycomb lattice of germane atoms with two sublattices A and B, as showed in Fig.1(a). Since the weak $\pi$ bond of germanene is not enough to stabilize plane structure, a kind of buckling structure appears for A/B sublattice. Similar to graphene, its lower energy electronic structure around  Fermi level is mainly dominated by two equivalent Dirac cone at $K$ and $K^{'}$, namely two valleys as shown in Fig.\ref{FIG1}(b). This equivalency is protected by the inversion symmetry of the A and B sublattices. If this symmetry is broken, valley-polarization will appear. Before investigating the system of TM adsorbed germanene, we perform the optimization calculations of the pristine germanene. The crystal constant and buckling distance are respectively 4.02 \AA\ and 0.61 \AA, which is in good agreement with previous reports\cite{cry1,cry2}.\\
\begin{figure}
  \includegraphics[width=3.5in]{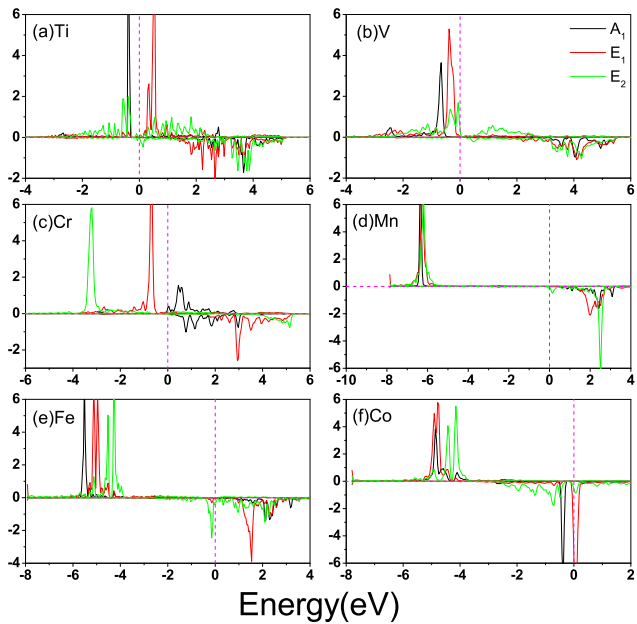}\\
  \caption{(Color online)Projected density of states for TM-germanene system. The positive and negative values denote spin-up and spin-down channels, respectively. The Fermi energy is set to zero.}\label{FIG4}
\end{figure}
\indent Then we study the TM adsorbed germanene system. In present work, we consider Ti, V, Cr, Mn, Fe, and Co as the TM adsorbates. Since the adsorption of Ni atom does not induce valley-polarization in Germanene, which is essential topic in this paper, we do not consider it in present work. Previous report by Wehling et al. \cite{sun1} indicated that electronic structures, adsorption geometry, and magnetic state of the 3\emph{d}-TM adsorbed graphene are very sensitive to the treatment of the local Coulomb interactions U of the TM \emph{d} orbital. Moreover, the first-principles Hubbard U is important to accurately describe the electronic structure of 3\emph{d} TM absorbed germanene due to the correlation of 3\emph{d} electrons is strongly dependent on its occupation and surrounding. Using the linear response approach\cite{addu} we calculate the Hubbard U of all the six TM atoms adsorbed on germanene, the values of U are listed in Tab.\ref{tab1}. The calculated results of the six kinds of 3\emph{d} TM adsorbed germanenes based on the Hubbard U obtained above are also listed in the Tab.\ref{tab1}. For each TM adsorbate, three adsorption sites including hollow, bridge, and topB of germanene as shown in Fig.\ref{FIG1}(a) are considered. The adsorption energy indicates that the hollow absorption site is the most stable configuration for all the six TM-germanene systems. Therefore, our discussions below concentrate on this absorption configuration. The absorption energies for all TM-germanene systems range from 1.489 eV to 4.807 eV which is much larger than that of TM-graphene\cite{gr1} (less than 1.0 eV). The results indicate that 3\emph{d} TM atom adsorbed germanene shows strong stability. For V, Cr, and Fe, the absorption energy agrees well with the results of uniform Hubbard U=4 eV\cite{gecal1}, and the difference is only around 0.1 eV. However, the adsorption energy of Ti, Mn and Co is 0.35, 1.43, and 0.93 eV larger than that of uniform Hubbard U=4 eV\cite{gecal1}, respectively. This difference mainly attributes to relatively larger Hubbard U in our calculations. Although the comparison of the adsorption energy between different Hubbard U is not that much meaningful, our recent work\cite{sun2} indicates that appropriate on-site Hubbard U can even influence the spin-polarized ground state. The calculated magnetic moments based on the Hubbard U derived from the linear response approach\cite{addu} show great difference from previous report with uniform Hubbard U\cite{gecal1}. The difference derives from the different electron occupation in the \emph{d} orbital which is sensitive to the on-site Hubbard U. For instance, when the Hubbard U=4.666eV the electrons of Ti-germanene almost averagely occupy the majority spin of $d_{xy},d_{x_2-y_2}$ and $d_{3z^2-r^2}$. Therefore the total magnetic moment is close to 3 $\mu_B$. However, when we take the Hubbard U=4.0eV the valence electrons fill only in the majority spin of $d_{xy}$ and $d_{xz}$ resulting in the magnetic moment close to 2 $\mu_B$.\\
\begin{figure}
  \includegraphics[width=3.2in]{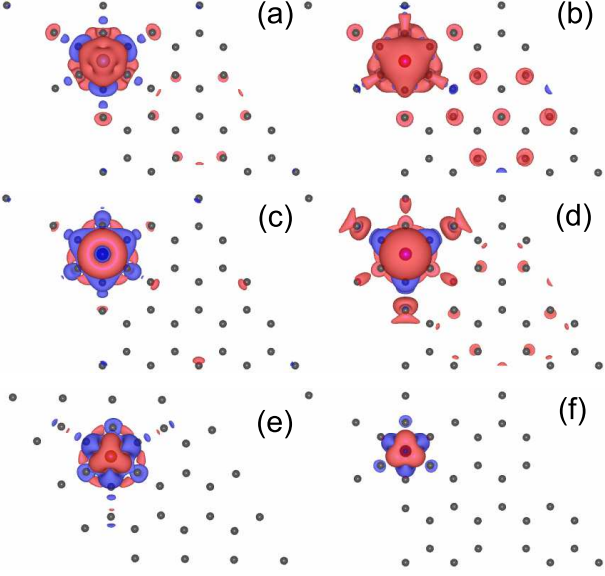}\\
  \caption{(Color online) Spin charge density distribution of (a)Ti-germanene (b)V-germanene (c)Cr-germanene (d)Mn-germanene (e)Fe-germanene (f)Co-germanene}\label{FIG3}
\end{figure}
\begin{figure}
  \includegraphics[width=3.0in]{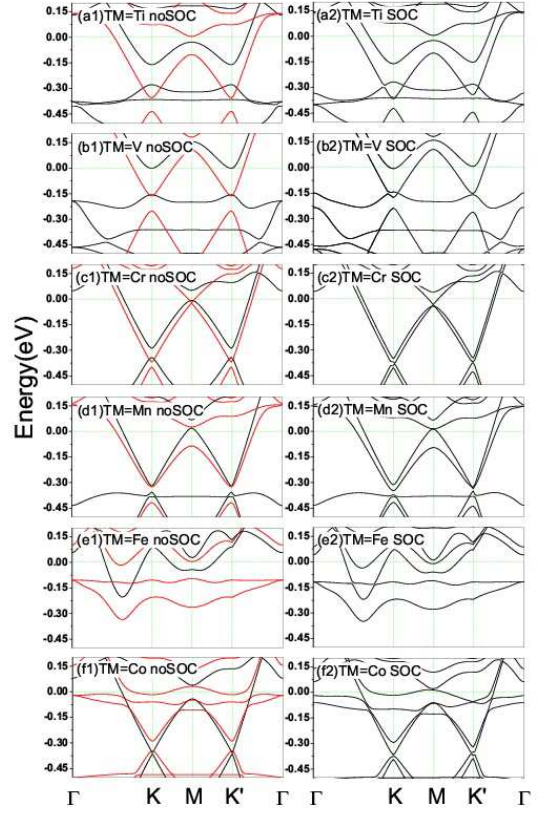}\\
  \caption{(Color online) Left column, (a1) - (f1): the calculated band structures for TM-germanene without SOC. The red and black curves represent the spin-up and spin-down channels, respectively. Right column, (a2) - (f2): the calculated band structures for TM-germanene with SOC. }\label{FIG2}
\end{figure}
\indent As mentioned above, pristine germanene consists of A and B sublattices with different height in the buckled configuration. The foreign TM adatom shows different distance with the nearest neighbor (NN) two sublattices d$(TM-Ge_A)$ and d$(TM-Ge_b)$, as listed in Table \ref{tab1}. The difference of these two distances is around 0.2-0.5 \AA. Thus, a local staggered AB-sublattice potential will be induced by the TM adatom. This effect is similar to vertical electric field applied to a single layer silicene\cite{electricSi}. The two valley $K$ and $K^{'}$ become un-dependent after the TM adsorption. The charge transfer $T_e$ between TM and germanene as shown in Tab.\ref{tab1} indicates that all TM-germanenes are n-type doped by the TM atom. The charge transfer decreases from Ti(1.153e) to Co(0.285e) attributing to the increase in the electronegativity from Ti to Co. The n-type doping of TM on germanene will be further discussed below.\\
\begin{figure}[ht!]
 \includegraphics[width=3.0in]{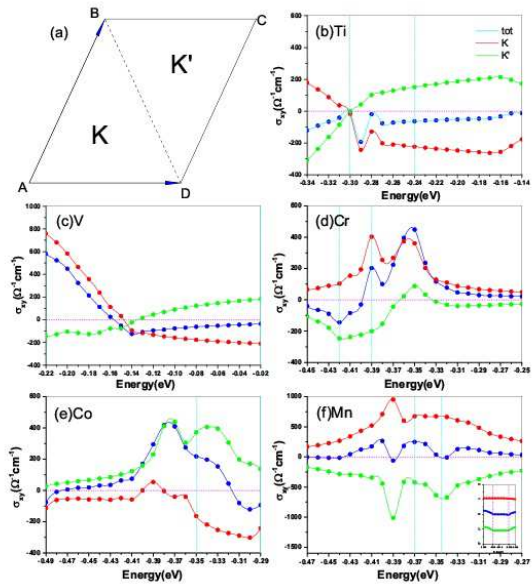}\\
  \caption{(Color online)(a)The first Brillouin zone for 4 $\times$ 4 germanene. AHC around the Dirac cone for (b)Ti-germanene, (c)V-germanene, (d)Cr-germanene, (e)Mn-germanene, and (f)Co-germanene. The Fermi energy is set to zero. The vertical dotted line represent the energy that we plot the Berry curvature distribution in Fig.6. The inset of (f) represents close inspect the AHC for Mn-germanene near the energy of -0.345. }\label{FIG5}
\end{figure}\\
\indent To understand the magnetic properties of TM-germanene as listed in Tab.\ref{tab1}, we plot the partial density of states (PDOS) of TM-germanene system without SOC in Fig.2. Although the $C_{3v}$ symmetry of the hollow site of pristine germanene is broken by the TM adsorption, the symmetry around the hollow site is close to $C_{3v}$. Therefore, we can qualitatively split the 3\emph{d} sub-shell of adatom into three group: $A_1$ symmetry group only included 3$d_{3z^2-r^2}$ state; the twofold degenerate $E_1$ group consists of $3d_{xz}$ and $3d_{yz}$; $3d_{xy}$ and $3d_{x^2-y^2}$ make up the $E_2$ group. For the case of Ti-germanene, the strong spin splitting almost makes the minority states totally locate in the conduction band, and the occupied double degenerate majority $E_2$ and majority $A_1$ give rise to 2.67 $\mu_B$ magnetic moment in the system which is nearly two times larger than that of uniform Hubbard U\cite{gecal1}. The spin charge density (SCD, defined as the difference between spin-up and spin-down charge density) as shown in Fig.3(a) indicates that the Ti shows anti-ferromagnetic coupling with its NN A site Ge and ferromagnetic coupling with its NN B site Ge. It is worthy to mention that, different from the extended distribution $E_1$ and $E_2$, the PDOS of dumbbell-shaped $A_1$ is very localized, indicating less interaction with neighboring Ge atoms. For V-germanene, the PDOS shares similar characteristics with those of Ti-germanene. The major difference is that the majority $E_1$ of V shows distribution below the Fermi level. The magnetic moment of V-germanene is close to 5.0 $\mu_B$(4.67$\mu_B$). The SCD as shown in Fig.3(b) indicates that the V shows ferromagnetic coupling with both its NN A and B site Ge. For Cr-germanene, some sizable Ge 4\emph{s}4\emph{p}-minority states distribute in the energy window of (-2.0,-3.0) implying a kind of antiferromagnetic coupling between Cr and Ge atoms, which lowers the total magnetic moment of Cr-germanene to 2.84 $\mu_B$ in the unit cell. The magnetic moment is smaller than previous report(4.0 $\mu_B$)\cite{gecal1}. The SCD as shown in Fig.3(c) indicates that the Cr shows anti-ferromagnetic coupling with both its NN A and B site Ge. The magnetic moment of Mn-germanene shows amazingly 5.75 $\mu_B$. After carefully examined the total DOS and partial DOS, we find that majority $E_1$, $E_2$, and $A_1$ states of Mn are fully occupied below the Fermi level resulting in the magnetic moment of the Mn close to 5.00 $\mu_B$. The fully occupied $A_1$ state is clearly shown in the Fig.3(d), which is perpendicular to the germanene plane. The SCD as shown in the Fig.3(d) indicates that Mn shows anti-ferromagnetic coupling with its NN A site Ge and ferromagnetic coupling with its NN B site Ge. Moreover, the next nearest neighbor (NNN) A and B site Ge atoms also show spin polarization same to the Mn adatom producing 5.75 $\mu_B$ of the system. In the case of Fe-germanene, the magnetic moment of Fe adatom is close to 3.9 $\mu_B$ because its majority states are fully occupied and only small fraction of minority $E_2$ and $E_1$ states are occupied just below the Fermi level, as shown in Fig.2(e). However, the SCD as shown in Fig.3(e) indicates that the Fe is anti-ferromagnetic coupling with both its NN A and B site Ge, reducing the magnetic moment of the Fe-germanene to 3.04 $\mu_B$. For the last case of Co-germanene, the majority states of TM are full occupied. But minority states $A_1$ and $E_2$ are also partially occupied locating below the Fermi level. The SCD as shown in Fig.3(f) indicates that the Co is anti-ferromagnetic coupling with both its NN A and B site Ge. The total magnetic moment of Co-germanene is 2.03 $\mu_B$. It is worth to mention that because of the 4\emph{s} states of 3\emph{d} TM is extended, the electron of 4\emph{s} easily transfers to 3\emph{d} orbits or 3\emph{s}3\emph{p} orbits of Ge atoms. Such electron transfer also affect the magnetic performance of TM adatom on germanene.\\
\indent Now we concentrate on the energy band structures of TM-germanene without/with SOC. As is showed in Fig.4(a1)-(f1), we find that the Dirac cones approximately remain except for Fe-germanene. Since the paper focuses on the valley-polarized phenomenon, Fe-germanene is excluded in later discussions. The Dirac cones of all TM-germanenes shift below the Fermi level due to the electron transfer from TM atom to germanene. Spin-splitting appears around the Dirac cone after the TM atom adsorbed on the hollow site of germanene, which is the prerequisites of AHE around the Dirac cone. The energy band structures with SOC are showed in Fig.4(a2)-(f2). Valley-polarized phenomenon is very evident for all TM-germanene systems due to the energy spectrum around Dirac cone exhibiting different characteristics between valley $K$ and $K^{'}$. For Ti-germanene and V-germanene, some states other than $K$ and $K^{'}$ appear around Dirac cones and cross with the two valleys. Together with the results of PDOS, we confirm that they mainly come from \emph{d} orbits of TM. These localized states will ruin the valley-polarized AHE of Ti/V-germanene, which is proved by Berry curvature calculations below. In comparison with other TM-germanene, we find that the Dirac cones of Cr-germanene and Co-germanene remain very well. After considered SOC, the $K$ of Cr-germanene and the $K^{'}$ of Co-germanene show energy band cross feature. For Mn-germanene, a local state derived from the \emph{d} states of Mn crosses through both valleys. Interestingly, there is an narrow gap around the two Dirac cones for Mn-germanene. The narrow energy gap is hoping to induce the system show quantum valley Hall effect, the details will be discussed below.\\
\begin{figure}[ht!]
  \includegraphics[width=3.0in]{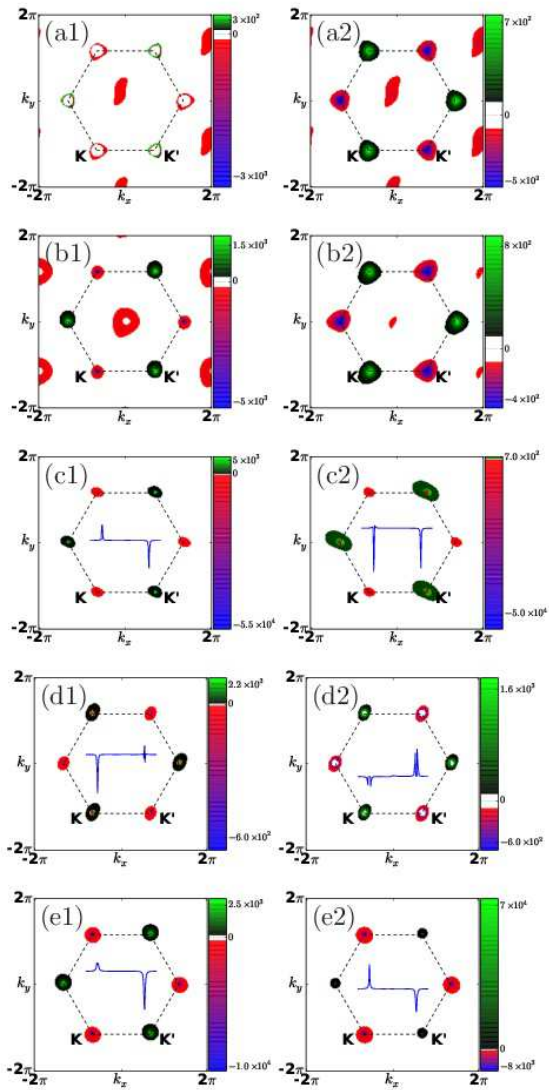}
  \caption{(Color online)The distribution of the Berry curvature in momentum space for Ti-germanene in the energy of (a1) -0.3 and (a2) -0.24,
  V-germanene in the energy of (b1) -0.22 and (b2) -0.02, Cr-germanene in the energy of (c1) -0.39 and (c2) -0.35, Co-germanene in the energy of (d1) -0.35 and (d2) -0.29, Mn-germanene in the energy of (e1) -0.37 and (e2) -0.345, the inset is one dimensional Berry curvature when $k_y = 0$.}\label{FIG6}
\end{figure}\\
\begin{figure}
 \includegraphics[width=3.0in]{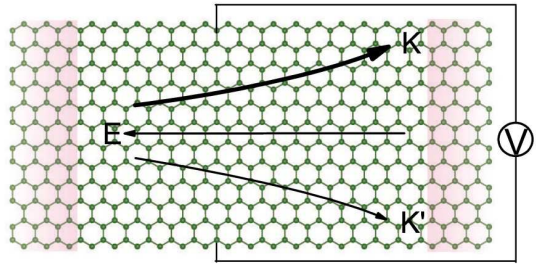}\\
\caption{An in-plane electric field will produce a transverse valley voltage between the two edge. Arrow E represent the direction of electric field, arrow $K$ and $K^{'}$ express the direction of electronic movement for different valley. The different of valley's AHC is showed by different weighted arrows.}\label{FIG7}
\end{figure}
\indent High symmetry point $K$ and $K^{'}$ locate on the opposite side of the high symmetric line BD of the first Brillouin zone of $4 \times 4 $ germanene as shown in Fig.5(a). When we integrate Berry curvature for one valley, the integration can be done only on one side of the line BD, and the integration accuracy of Wannier interpolation have been proved by previous works\cite{qahe1,qahe2}. For Ti-germanene, an interesting phenomenon appears from the result of berry curvature that the sign of its Berry curvature from valley $K$ and $K^{'}$ is reversal if we shift the Fermi level around -0.30eV, as showed in Fig.5(b). The reversion of the Berry curvature will produce the reversion of the voltage derived from VHE. However, the PDOS of Fig.2(a) shows that strong localized $A_1$ state distributes within the energy window of (-0.45eV, -0.3eV). The $A_1$ state localization destroys the good quantum number of the valleys, namely Berry curvature has finite value contributed by the k-points other than $K$ and $K^{'}$, as shown in Figs.6(a1) and (a2). Such Berry curvature other than $K$ and $K^{'}$ will cover the contribution of the valley $K$/$K^{'}$ to some extent. The system although may show the voltage similar to VHE, the contribution is hardly to be distinguished, which is unsuitable for QVHE. The system V-germanene shares the similar feature with the Ti-germanene as shown in Figs.6(b1) and (b2). We should note that, although there are contribution from the k-points other than the valley $K$/$K^{'}$ for Ti-germanene and V-germanene, the main contribution to the Berry curvature still comes from the two valleys. We predict that the Ti-germanene and V-germanene systems are also hoping to exhibit VHE voltage in experiments although it is not purely contributed by the valley polarization.\\
\indent The Berry curvatures of Cr-germanene when the Fermi energy is -0.39 eV and -0.35 eV respectively are shown in Fig.6(c1) and (c2). The AHC with valley-polarized characteristics of Cr-germanene in function of the Fermi level around the Dirac cone are shown in Fig.5(b). The solid lines in the figure are obtained by fitting scatter data with cubic spline functions. The anomalous Hall conductivities from half Brillouin zone integration almost totally come from the valley $K$ and $K^{'}$. AHC contributed by valley $K$ keeps positive value. However, the AHC derived from the $K^{'}$ change from negative to positive within the energy window (-0.36eV, -0.33eV). The results indicate that the valley current derived from $K^{'}$ is reversal in the energy window. According to the energy band as shown in Fig.3 and Berry curvature distribution as depicted in Fig.6(d1) and (d2), such reversion is closely related to the cross of Fermi level with special energy band at $K^{'}$, which make the Berry curvature around the center zone of $K^{'}$ change its sign. But no essential change is happened for $K$ when Fermi level shift from -0.39eV to -0.35eV. Interestingly, when the Fermi level is larger than -0.4 eV, the contribution to the AHC derived from $K$ is larger than that of $K^{'}$ producing the total AHC change from negative to positive. When the Fermi level shifts to -0.35, the AHC reaches its maximum. The above results indicate that the AHC can be reversed through shifting the Fermi level, which will produce the reversion of the anomalous Hall voltage (AHV). Such feature of Cr-germanene is very useful for the information devices based on AHE. Similar feature also exhibits in the system Co-germanene. The anomalous Hall conductivities from half Brillouin zone integration almost totally come from the valley $K$ and $K^{'}$. AHC contributed by valley $K^{'}$ keeps positive value. However, the AHC derived from the $K$ changes from negative to positive when the Fermi level is around -0.39eV. However, the values are very small producing the total AHC reach its maximum around the energy. According to the shift of energy band as shown in Fig.3 and Berry curvature distribution as depicted in the Fig.6(e1) and (e2), such reversion is closely related to the cross of Fermi level with special energy band at $K$, which make the Berry curvature around the center zone of $K$ changes its sign. But no essential change is happened for $K^{'}$. And then, along with the increase in the Fermi level, the AHC derived from $K$ increases and reaches its maximum around -0.30eV. When the Fermi level is larger than -0.30eV, the total AHC changes from positive to negative. The results indicate that the valley current derived from $K$ is reversal similar to Cr-germanene. Actually, when the Fermi level is smaller than -0.48eV the total AHC of Cr-germanene is also negative.\\
\indent The AHC of Mn-germanene is shown in Fig.5(f). The values of AHC derived from $K$ and $K^{'}$ remain positive and negative, respectively, within the energy window considered in present work. Moreover, the evolution of the AHC derived from $K$ and $K^{'}$ with the increase in the Fermi level approximately shows mirror symmetry resulting in the total AHC of the system always slight. However, we find that quantum valley hall effect appears for this system, as showed in the inset of Fig.5(f). When the Fermi level is around -0.35 eV, quantized Hall conductance platform appear for $K$ and $K^{'}$ and the value of $C(K) \approx -C(K^{'}) \approx 1$. The width of the platform is about 10 meV, which corresponds to the band gap around the Dirac cone.\\
\indent Based on the above results, the AHE in TM-germamene can be modulated by shifting Fermi level of the systems. Previous work\cite{siBN} has reported that when silicene and BN form superlattice, the electron transfer from silicene to BN leading to the up-shift of Dirac Cone of silicene. We predict that if the TM-germanene grows on a chemical inertia substrate such as BN, the electron transfer between TM-germanene and the substrate will shift the Fermi level of TM-germanene and at the same time maintain its property of Dirac cone. Moreover, it has been reported that the Fermi level of two dimensional systems, such as single/bilayer graphene and graphene-$MoS_2$ heterojunction, can be tuned just by applying gated voltage in the direction perpendicular to the two dimensional plane\cite{fermi1,fermi2,fermi3}. Therefore, we expect that a z-direction gated voltage would be excellent method to tune the Fermi level for our TM-germanene system grown on chemical inertia substrate.\\
\indent In general valley Hall material, such as $MoS_2$ and graphene, equal amounts of Hall current from each valley flow in opposite directions due to time reversal symmetry, so that no net Hall voltage is produced. In TM-germanene system, this can be easily realized. The mechanism is depicted in Fig.7. When an in-plane electric field is applied, the electron of different valley acquires opposite anomalous velocity proportional to the Berry curvature in the transverse direction. If the Fermi level shifts to specific energy, unequal Berry curvature between two valley ultimately leads to a valley Hall voltage between two boundary. A net valley-polarized electric current can be acquired in the longitudinal direction. Moreover, the AHV of TM-germanene, such as Cr-germanene and Co-germanene, can be reversed by shifting the Fermi level.\\
\section{Conclusion}
\indent On the basis of the first-principles calculations, we report the structure and electronic properties of 3\emph{d} TM adsorbed germanene. Rely on the comparison with previous reports, we find that a proper Hubbard U is important to accurately describe the properties of TM-germanene systems. The Berry curvature calculations indicate that valley-polarized AHE can be realized in the Cr, Mn, or Co adsorbed systems. Furthermore, this kind of valley-polarized AHE can be effectively modulated by shifting the Fermi level of the systems.\\
\begin{acknowledgments}
This work is supported by the National Natural Science Foundation of China (Grant Nos. 10874143 and 10974166), the Program for New Century Excellent Talents in University (Grant No. NCET-10-0169), and the Scientific Research Fund of Hunan Provincial Education Department (Grant Nos. 10K065, 10A118, 09K033).
\end{acknowledgments}


\end{document}